\documentclass[aps,physrev,preprint,groupedaddress]{revtex4-2}
\usepackage{graphicx} 
\usepackage{amsmath}
\usepackage{subcaption} 
\usepackage[scaled=0.92]{helvet} 

\DeclareCaptionFont{subfiglabel}{%
  \sffamily                  
  \bfseries                  
  \fontsize{10}{10}\selectfont 
}

\newcommand{\figref}[2][]{%
\ref{#2}%
  \if\relax\detokenize{#1}\relax
  \else
    #1%
  \fi
  }

\begin{document}

\title{Bipotentiostatic Control Unlocks Flashing Ratchet Features in Ion Pumps}

\author{Eden Grossman}
\affiliation{School of Electrical Engineering, Tel Aviv University, Tel Aviv 6997801, Israel}

\author{Alon Herman}
\affiliation{School of Electrical Engineering, Tel Aviv University, Tel Aviv 6997801, Israel}

\author{Keren Shushan Alshochat}
\affiliation{School of Electrical Engineering, Tel Aviv University, Tel Aviv 6997801, Israel}

\author{Dafna Amichay}
\affiliation{School of Electrical Engineering, Tel Aviv University, Tel Aviv 6997801, Israel}

\author{Ilan Bijaoui}
\affiliation{School of Electrical Engineering, Tel Aviv University, Tel Aviv 6997801, Israel}

\author{Gideon Segev\footnote{gideons1@tauex.tau.ac.il}}
\affiliation{School of Electrical Engineering, Tel Aviv University, Tel Aviv 6997801, Israel}

\date{\today}

\begin{abstract}
The selective separation of same-charge ions is a longstanding challenge in resource recovery, battery recycling, and water treatment. Theoretical studies have shown that ratchet-based ion pumps (RBIPs)  can separate ions with the same charge and valance by driving them in opposite directions according to their diffusion coefficients. This process relies on frequency dependent current reversal, a unique feature of ratchets in which the particle current direction is inverted with the input signal frequency. Previous experimental demonstrations of RBIPs achieved ion pumping against electrostatic forces and water deionization, but lacked frequency-dependent current reversal and control of the asymmetry of the device. Here, we report the first experimental realization of these key functionalities by driving RBIPs with a bipotentiostat. Complementary input signals applied to RBIP contacts unlock a flashing ratchet-like behavior, and enhances the device performance by an order of magnitude compared to the prior floating-drive approach. The enhanced control of the electrostatic potential at the RBIP surfaces leads to frequency dependent current reversals, and the addition of a potential offset to the input signal enables tuning the amplitude asymmetry of the device. This flashing ratchet functionality provides a significant step towards the realization of ratchet driven selective ion separation systems.
\end{abstract}


\maketitle
\section{Introduction}
Selective ion separation is a major challenge in water treatment, resource recovery, battery recycling, and other applications \cite{Sholl2016SevenWorld,Werber2016MaterialsMembranes}. One of the leading research directions for achieving ion-ion selectivity is filtration with sub-nanometer pore membranes. In these membranes, ion-ion selectivity is governed by differences in ion hydration energies and transport properties through the membrane pores \cite{Tang2020IonMembranes,Epsztein2020TowardsPores,Kingsbury2025ATransition}. Although such membranes have shown promise, they demonstrated limited selectivity for ions with the same charge and valence and face significant scalability challenges \cite{Tang2020IonMembranes, Epsztein2020TowardsPores}. Furthermore, since the separation mechanism is determined by the specific properties of each ion, distinct separation processes require the development of different membranes.  

Ratchets have been widely studied for particle separation. Ratchets are devices that utilize a spatially asymmetric, time-modulated force to drive a steady-state particle flux. A unique feature of ratchets is current reversal, in which the particle flow direction is inverted with frequency. Theoretical studies have shown that since the current reversal frequency (also termed the stopping frequency) is determined by the particle properties, ratchets can separate particles by driving them in opposite directions according to their mass, size, and mobility \cite{Hanggi2009ArtificialNanoscale, Sapik2019TunableMobility, Tarlie1998OptimalForce, Kettner2000DriftRatchet, Sapik2019TunableMobility}. Nano-particle current reversal and separation by shape was experimentally demonstrated in nano-structured nanofluidic devices in which an electric signal was applied between two  electrodes positioned on opposite sides of the device, resulting in a tilting ratchet functionality \cite{Skaug2018NanofluidicMotors, Schwemmer2018ExperimentalMotor}. Since ratchets can drive particles with different properties in opposite directions, they are uniquely suited for selective separation, even at trace target particle concentrations. Furthermore, as the particle sorting properties are determined by an applied external force, separation processes can be tuned in real time, and the same device can be modified relatively easily for various separation processes.

Flashing ratchets are ratchets in which the potential distribution through the device, $U(x,t)$, is a product of  the spatial potential distribution, $ V(x)$,  and a temporal modulating function, $g(t)$, such that $U(x,t) = V(x)g(t)$ \cite{Kedem2017HowCurrent, Lau2017AnBiology}. Theory shows that flashing ratchets drive particles at the highest velocities when operated with square wave input signals \cite{Tarlie1998OptimalForce}:
\begin{equation}\label{eq:g_def}
g(t) \;=\; \begin{cases}
1, & 0 < t < d_{c} \cdot T \\
\alpha, & d_{c} \cdot T < t < T
\end{cases} \\
\end{equation}
where $T$ is the input signal temporal period, $d_c$ is its duty cycle and $\alpha$ is the amplitude asymmetry coefficient. Theoretical analyses have demonstrated that flashing ratchets can drive ions of identical charge and valence in opposite directions according to their diffusion coefficients, as well as to drive both cations and anions in the same direction resulting in ambipolar ion transport \cite{Herman2023Ratchet-BasedSeparations, Herman2024AmbipolarMembranes}. Device simulations have shown that current reversal with frequency, which is essential for this separation process, is possible only for $-1<\alpha<0$. Thus, controlling $\alpha$ and demonstrating current reversals are essential prerequisites for ratchet-driven ion separation.

We have recently realized first-of-their-kind ratchet-based ion pumps (RBIPs) \cite{Kautz2025ASeparations}. RBIPs were fabricated by coating the two surfaces of anodized aluminum oxide (AAO) wafers with metals. Then, RBIPs were placed as active membranes between two electrolyte compartments. When alternating input signals were applied between the two RBIP surfaces, the devices were shown to drive ions against an electrostatic force, deionize water in an electrodialysis cell, and tune the onset of electrochemical reactions \cite{Kautz2025ASeparations, Amichay2025TuningPumps}.  These experimental demonstrations utilized a "floating drive" configuration, in which the input signal imposed an electrostatic potential difference between the two RBIP surfaces with no well-defined reference potential. As a result, the electrostatic potential difference between each RBIP contact and adjacent bulk solution followed a capacitive charging-discharging behavior, generating a time-averaged driving force via the nonlinear capacitance of the electric double layers \cite{Kautz2025ASeparations}. Thus, although demonstrating a ratchet functionality, the floating drive configuration cannot control the amplitude asymmetry coefficient, $\alpha$, and has not shown clear frequency induced current inversions. Consequently, these efforts lack key flashing-ratchet characteristics which are essential for highly selective ratchet-driven ion separation.

In this work, we demonstrate that driving RBIPs with a bipotentiostat configuration results in a flashing ratchet-like functionality. We show that the amplitude asymmetry coefficient, $\alpha$,  can be tuned by adding a constant potential bias to the RBIP contacts.  Frequency induced reversal of the current and voltage measured between two Ag/AgCl wires was also demonstrated. Furthermore, this potential control scheme yields a performance enhancement of up to 1000\% compared to the floating RBIP drive. The combination of enhanced output and the realization of key features characteristic of flashing ratchets may open new pathways toward the development of ratchet-driven selective ion separation technologies.

\section{Methods} \label{methods section}
RBIPs were fabricated from nanoporous anodized aluminum oxide (AAO) wafers (InRedox Materials Innovation) with pore diameters of 40 nm and thickness of 50 $\mu$m. Prior to metal deposition, the wafers were annealed in air at 650 °C for approximately 10 h. A 40 nm (planar equivalent) gold thin film was then deposited on both wafer surfaces using magnetron sputtering (Vinci Technologies PVD 20), forming the RBIP contacts without blocking the pores. The devices were placed as membranes separating two reservoirs in an electrochemical cell, with both reservoirs filled with 1 mM KCl aqueous solution in all experiments. More details on the RBIP fabrication process and morphology can be found in prior work \cite{Kautz2025ASeparations}.

\begin{figure*}[htbp!]
\setcounter{subfigure}{0}
\includegraphics[width=17.2cm]{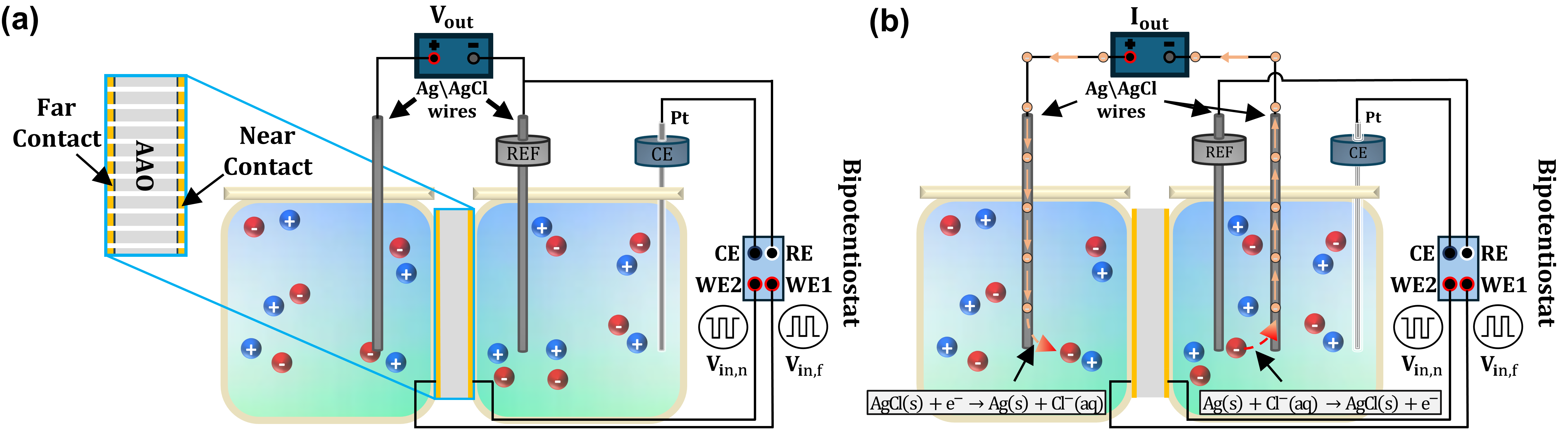}
\caption{A schematic illustration of the experimental system: (a) $V_{out} $ measurement configuration. (b) $I_{out}$ measurement configuration}
\label{fig: simple system}
\end{figure*}

Figure \ref{fig: simple system} shows a schematic illustration of the experimental setup.  The electrostatic potential of the RBIP gold surfaces was alternated using a BioLogic VSP300 bipotentiostat. Each gold surface was electrically connected to the working electrode lead of a separate bipotentiostat channel. The working electrode potential (vs. reference) of each of the bipotentiostat channels was configured to follow the output of a Keysight 33522B waveform generator. Both channels shared a platinum wire counter electrode and an Ag/AgCl wire quasi-reference electrode. More details on the electrical connections used to apply the electric signals to the RBIP can be found in appendix \ref{Electrical configuration}. 

Hereafter, we refer to the gold surface adjacent to the compartment containing the counter and reference electrodes as the near contact, to which the input signal $V_{in,n}$ is applied. The opposite gold surface is referred to as the far contact, and the corresponding input signal is noted $V_{in,f}$ (Figure \ref{fig: simple system}). The input signals applied to the two contacts are periodic rectangular waves with an identical frequency, $f=1/T$:
\begin{equation}\label{eq:vin_definitions}
\begin{aligned}
V_{in,f}(t) &=
\begin{cases}
V_{offset,f} + \dfrac{V_{p-p}}{2}, & 0 < t < d_{c,f} \cdot T \\
V_{offset,f} - \dfrac{V_{p-p}}{2}, & d_{c,f} \cdot T < t < T
\end{cases} \\
V_{in,n}(t) &=
\begin{cases}
V_{offset,n} + \dfrac{V_{p-p}}{2}, & 0 < t + \Delta t < d_{c,n} \cdot T \\
V_{offset,n} - \dfrac{V_{p-p}}{2}, & d_{c,n} \cdot T < t + \Delta t < T
\end{cases}
\end{aligned}
\end{equation}
$\Delta t= \frac{\theta}{360^\circ}\,T$  is the time delay of $V_{in,f}$ relative to $V_{in,n}$, and  $\theta$ is the corresponding phase in degrees. $V_{offset,f}$ and $V_{offset,n}$ are the voltage offsets of each signal relative to the reference electrode, $d_{c,f}$ and $d_{c,n}$ are the duty cycles of the two signals and $V_{p-p}$ is the peak-to-peak amplitude of the signals.

The device output was obtained by measuring the current or voltage between two Ag/AgCl wires placed across the RBIP using Keysight 34465A digital multimeters. Figures \figref[a]{fig: simple system} and \figref[b]{fig: simple system} show respectively schematic illustrations of the output voltage and current measurement setups.
The net ratchet-induced current and voltage,  $\Delta\overline{I}_{\text{out}}$  and $\Delta\overline{V}_{\text{out}}$ respectively, were calculated by subtracting the mean output measured when the ratchet was OFF ($V_{\text{in,f}} = V_{\text{in,n}} = 0$V) from the mean output measured when the ratchet was ON. More details on the definition of the output current and voltage can be found in appendix \ref{RBIP performance characterization}. All reported currents were normalized by the active area of the membrane, $A=0.159\space cm^2$, and are presented in units of $\mu A/cm^2$.

\section{Results}
\subsection{Bipotentiostat driven RBIP}
\begin{figure}[htbp!]
    \centering
      \subfloat[]{\includegraphics[width=8cm]{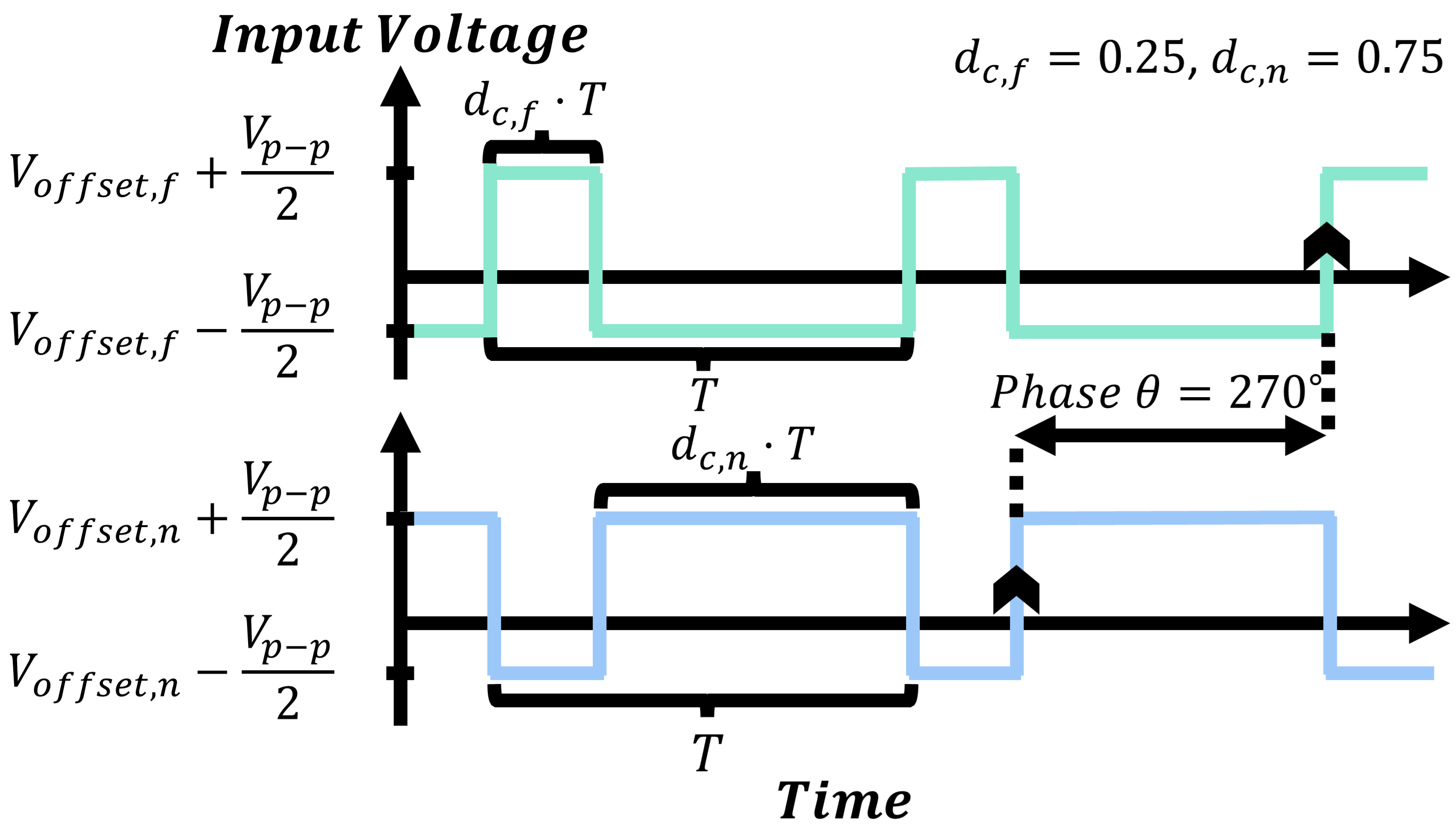}\label{fig: phase explained}}
    \hfill
    \subfloat[]{\includegraphics[width=8cm]{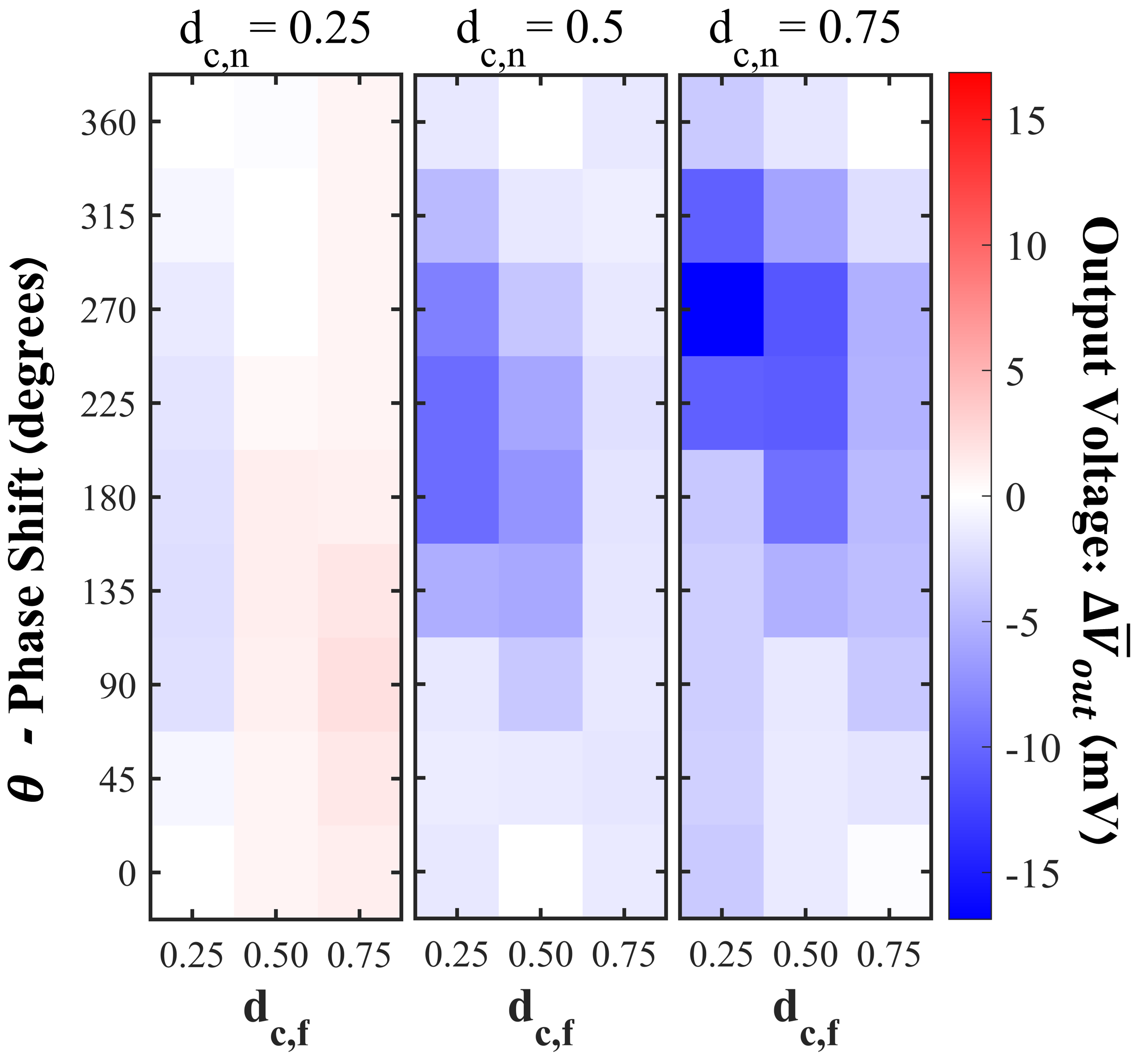}\label{fig: heatmap}}
  \caption{(a) An example for complementary $V_{in,f}(t)$ and $V_{in,n}(t)$ input signals with $d_{c,f}=0.25$, $d_{c,n}=0.75$, and phase shift $\theta=270^\circ$. (b) The RBIP output voltage as a function of $d_{c,f}$, $d_{c,n}$, and $\theta$, at a frequency of 100 Hz.}
  \label{fig:phase_Sweep}
\end{figure}

The RBIP performance was studied for input signals as defined in equation \ref{eq:vin_definitions}. Figure~\ref{fig: phase explained} shows an example of the two input signals and the main input signal parameters.
The potential applied to the RBIP contacts was maintained between $-0.05\mathrm{V}$ and $0.45\mathrm{V}$ versus the reference electrode in all experiments. Cyclic voltammetry confirmed that no electrochemical reactions occurred on the RBIP gold surfaces within this range (see Appendix~\ref{Working range}). Unless specified otherwise, the offset voltages were set to \( V_{\mathrm{offset,f}} = V_{\mathrm{offset,n}} = 0.2\,\mathrm{V} \) and the peak-to-peak voltage amplitude was set to \( V_{p-p} = 0.5\,\mathrm{V} \).

Figure~\ref{fig: heatmap} shows the device output voltage as a function of the input signals duty cycle ($d_{c,f}$, $d_{c,n}$) and phase shift ($\theta$). The input signals frequency was 100 Hz.
The device generated both positive and negative outputs. The highest positive output ($\Delta\overline{ V}_{\text{out}}=2 mV$) was obtained with $d_{c,f} = 0.75$, $d_{c,n} = 0.25$, and $\theta = 90^\circ$. The most negative output ($\Delta\overline{ V}_{\text{out}}=-16.9 mV$) was obtained with $d_{c,f} = 0.25$, $d_{c,n} = 0.75$, and $\theta = 270^\circ$. For duty cycles of $d_{c,f}=d_{c,n} = 0.5$, the phase shift yielding the largest absolute output ($\Delta \overline{V}_{out}=$-7.1 mV) was 180°.
In all three cases, the optimal output was obtained under complementary input conditions: $d_{c,f}=1-d_{c,n} $, and $\theta=d_{c,n}\cdot360^\circ $ , i.e. the high state of one signal always coincides with the low state of the other. This implies that the two surfaces of the RBIP do not operate as two independent sub-devices, and operating them in concert results in a synergy that enhances the device output. 
Based on this analysis, the complementary input signals were used in all subsequent measurements (unless stated otherwise). For simplicity, in the following discussion, applied signals are defined by \( V_{in,f} \) (see figure \ref{fig: simple system}), and the duty cycle is defined according to \( d_c = d_{c,f} \). 

\begin{figure}[htbp!]
    \centering
    \includegraphics[width=8.6cm]{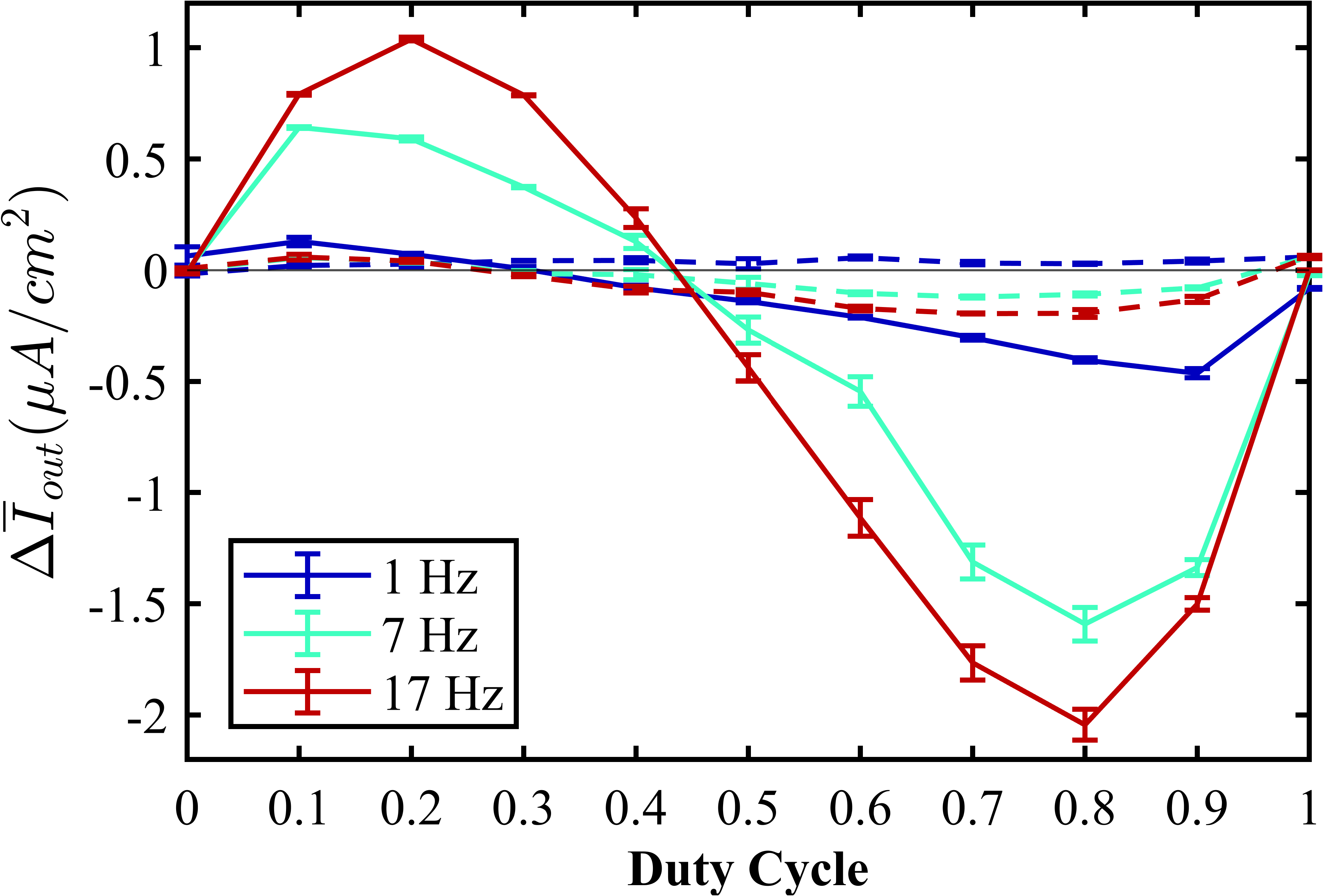}
    \caption{The RBIP output current as a function of the input signal duty cycle, when driven with the bipotentiostat  (solid lines) and with a floating drive (dashed lines).  Traces show the mean of duplicate measurements; error bars indicate maximum and minimum values.}
    \label{fig: VSP vs floating}
\end{figure}

Figure~\ref{fig: VSP vs floating} shows a comparison between the output current density obtained when the RBIP was driven with the bipotentiostat and with a floating drive. In both cases the input signal amplitude was $V_{p-p}= 0.5 V$. To account for possible sample degradation, the full parameter scan was conducted twice in each configuration alternating between the drives, resulting in two complete scans for each configuration.  Error bars indicate the maximum and minimum output obtained for each input signal, while traces show their mean values.   
As other ratcheting devices, the output is negligible when a constant bias is applied (duty cycle of 0 and 1). However, once an alternating input signal is applied, the device generates a significant output. The device generated substantially higher output currents when driven with the bipotentiostat, exceeding those of the floating drive by more than an order of magnitude. Moreover, the bipotentiostat enabled operation at frequencies where the floating drive failed to generate any measurable output. The corresponding peak current values are summarized in Table~\ref{tab: VSP vs floating}.

\begin{table}[htbp!]
\centering
\begin{tabular}{c|c|c}
\hline
Frequency & Bipotentiostat $\Delta\overline{I}{out}$ & Floating $\Delta\overline{I}{out}$ \\ 
\hline
17 Hz & $-2.04\,\mu A/cm^2$ ($d_c=0.8$) & $-0.20\,\mu A/cm^2$ ($d_c=0.7$) \\
7 Hz & $-1.59\,\mu A/cm^2$ ($d_c=0.8$) & $-0.12\,\mu A/cm^2$ ($d_c=0.7$) \\
1 Hz & $-0.46\,\mu A/cm^2$ ($d_c=0.9$) & No output (noise floor) \\ 
\hline
\end{tabular}
\caption{The largest output current obtained in the bipotentiostatic and floating drives in Figure \ref{fig: VSP vs floating}.}
\label{tab: VSP vs floating}
\end{table}

\subsection{Velocity Reversal and Ion Pumping}
Figures \figref[a]{fig:Parameter_sweep} and \figref[b]{fig:Parameter_sweep} show respectively the output voltage and current density as a function of the input signal duty cycle and frequency. As expected in ratchet devices, the output is low for very high and very low frequencies. These results are qualitatively similar to results obtained in flashing ratchets driving electrons \cite{Kedem2017HowCurrent}. The duty cycle and frequency affect the output magnitude and sign, with peak performance at 100 Hz and duty cycle 0.7: $\overline{I}{out}=-5.17,\mu A/cm^2$ and $\Delta\overline{V}{\text{out}}=-53.96 mV$. These values significantly exceed previously published results for experiments under similar conditions \cite{Kautz2025ASeparations}.

Figure \figref[c]{fig:Parameter_sweep} shows the output current and voltage as a function of the input signal frequency for a duty cycle of 0.1. Both the output current and voltage show a frequency induced reversal. Because of the very high reaction rate for $Cl^-$ on the Ag/AgCl electrodes, the voltage measured between the Ag/AgCl electrodes is essentially the electrochemical potential difference for  $Cl^-$ between the two compartments. Thus, a change in the sign of the measured voltage is indicative of a change in the direction of the driving force for  $Cl^-$ applied by the RBIP.  The current measured between the Ag/AgCl electrodes is the compensating charge flow required to maintain bulk solution electroneutrality when $K^+$ and $Cl^-$ are transported through the membrane at different fluxes \cite{Kautz2025ASeparations}. 
The electrodes enable this current by oxidizing $\text{Ag}(s)$ with $Cl^-$ in one compartment and reducing $AgCl$ in the other. A current sign change indicates that the difference in the $Cl^-$  and $K^+$ ion flux has reversed, crossing zero when the fluxes are equal. 
Since the AAO pores are positively charged at neutral pH, the ion flux through the RBIP will be dominated by the $Cl^-$  flow \cite{Kim2019StackedMembrane,Kim2018ElectricallyControl,Petukhov2017LiquidMembranes}. Therefore the observed change in current sign is most likely the reversal of the directional flow of $Cl^-$.
Frequency-dependent velocity reversal is a fundamental feature of flashing ratchet systems and enables high-precision selective ion separation \cite{Herman2023Ratchet-BasedSeparations}. Thus, experimentally demonstrating frequency reversal is a critical step toward realizing tunable, ratchet-driven selective ion separation systems. 

\begin{figure*}[htbp!]
  \includegraphics[width=17.2cm]{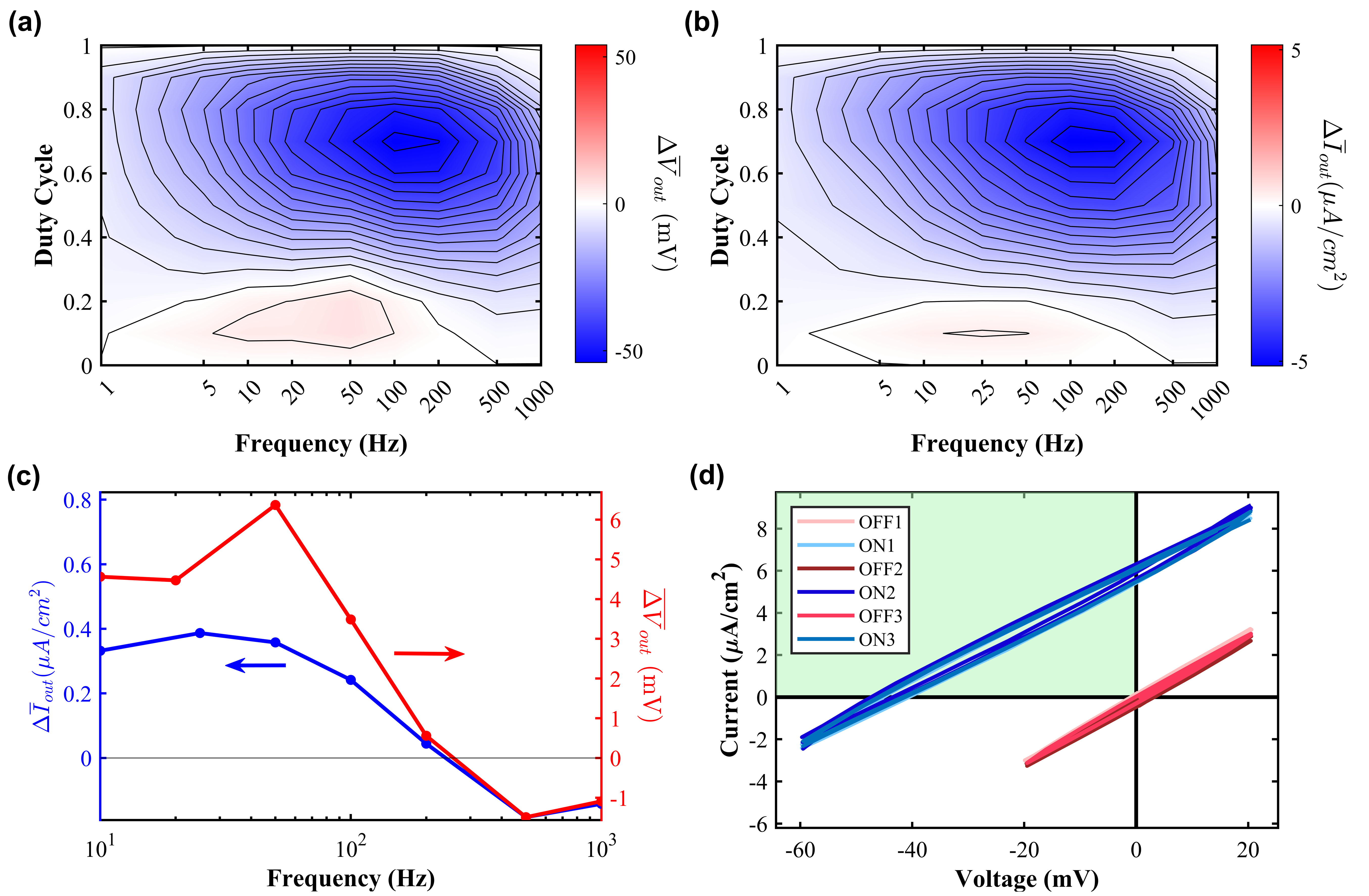}
  \caption{The RBIP output voltage (a) and current (b) as a function of the input signal duty cycle and frequency. (c) The output current density and voltage as a function of input signal frequency at $d_c=0.1$. (d) Cyclic voltammograms measured between the Ag/AgCl wires while the RBIP was operated and when it as OFF. During the ON state, a square wave signal was applied with $f = 100$ Hz, $d_c = 0.7$. In OFF1 and OFF2, the device contacts were not connected (open-circuit contacts). In OFF3, DC inputs of $V_{\text{in,f}} = 0.3$ V and $V_{\text{in,n}} = 0.1$ V were applied, corresponding to the time-averaged values of the ON signal. The chronological order of the measurements is as they appear in the legend.}
  \label{fig:Parameter_sweep}
\end{figure*}

Ions pumping against an electrostatic force was studied by analyzing cyclic voltammograms (CVs) of the device during the RBIP operation and while it was OFF. CV measurements were taken between the Ag/AgCl wires used for current measurements (Figure \figref[b]{fig: simple system}), with a separate potentiostat channel. Figure \figref[d]{fig:Parameter_sweep} shows the measured voltammograms. When the RBIP was ON, two complementary rectangular input signals with $f = 100$ Hz, and $d_c = 0.7$ were applied to the RBIP contacts. Voltammograms were also measured with with the RBIP disconnected (OFF1 and OFF2), and under a constant bias of $V_{\text{in,f}} = 0.3$ V and $V_{\text{in,n}} = 0.1$ V, corresponding to the time-averaged values of the signals applied to the RBIP (OFF 3).  The chronological order of the measurements is as they appear in the legend. In all cases, the CVs are linear  with a slope that is determined by the ohmic resistance of the reservoirs and transport through the membrane. As in prior works, no current rectification is observed, nevertheless the RBIP output is significantly higher than earlier report which implemented a floating drive \cite{Kautz2025ASeparations}.
Using the bipotentiostatic drive induces a voltage shift of approximately $\Delta\overline{V}{\text{out}}$ to the CV curve. As a result,  the CV curves pass through the second quadrant, where the applied voltage and the measured current are opposed. This demonstrates that the RBIP drives ions against an external electrostatic force, practically functioning as a time-averaged DC voltage source \cite{Roeling2011OrganicWork}.

\subsection{Bias-Induced Asymmetry Tuning}
 Most theoretical analyses of flashing ratchet assumed that the device potential alternates between  $V_+(x)=V(x)$ and $V_-(x)=\alpha V(x)$.\cite{Rozenbaum2019SymmetryRatchets, Kedem2017HowCurrent, Herman2023Ratchet-BasedSeparations}. Device modeling shows that a rectangular input signal with $\alpha = -1$ results in a particle velocity for duty cycle curve that is antisymmetric about $d_c=0.5$ \cite{Rozenbaum2019SymmetryRatchets}. As $\alpha$ increases from -1 to 0, the stopping duty cycle (the duty cycle where net velocity is reversed) shifts from $d_c=0.5$ toward $d_c=0$ or $d_c=1$\cite{Kedem2017HowCurrent, Herman2023Ratchet-BasedSeparations}.  

By assuming that the potential alternates with respect to 0 V these models disregard the potential of zero charge of the electrodes (or band bending at the metal-semiconductor interface in electron ratchets). However, the potential of zero charge for gold in KCl aqueous solution can be as high as 0.4 V vs. Ag/AgCl, and is highly sensitive to the crystalline structure and surface orientation \cite{Kasuya2016AnionMeasurement, Trasatti1971WorkFunctions, Hamelin1976TheFaces}. Hence, in experimental systems, the electrostatic potential at the electrodes (with respect to the electrostatic potential at the bulk solution) alternates between values that are determined by the potential of zero charge and the input signal amplitude and offset. Thus, an effective amplitude asymmetry coefficient can be tuned by varying the offset potential applied to the RBIP contacts. 

\begin{figure}[htbp!]
\centering
    \subfloat[]{\includegraphics[width=8cm]{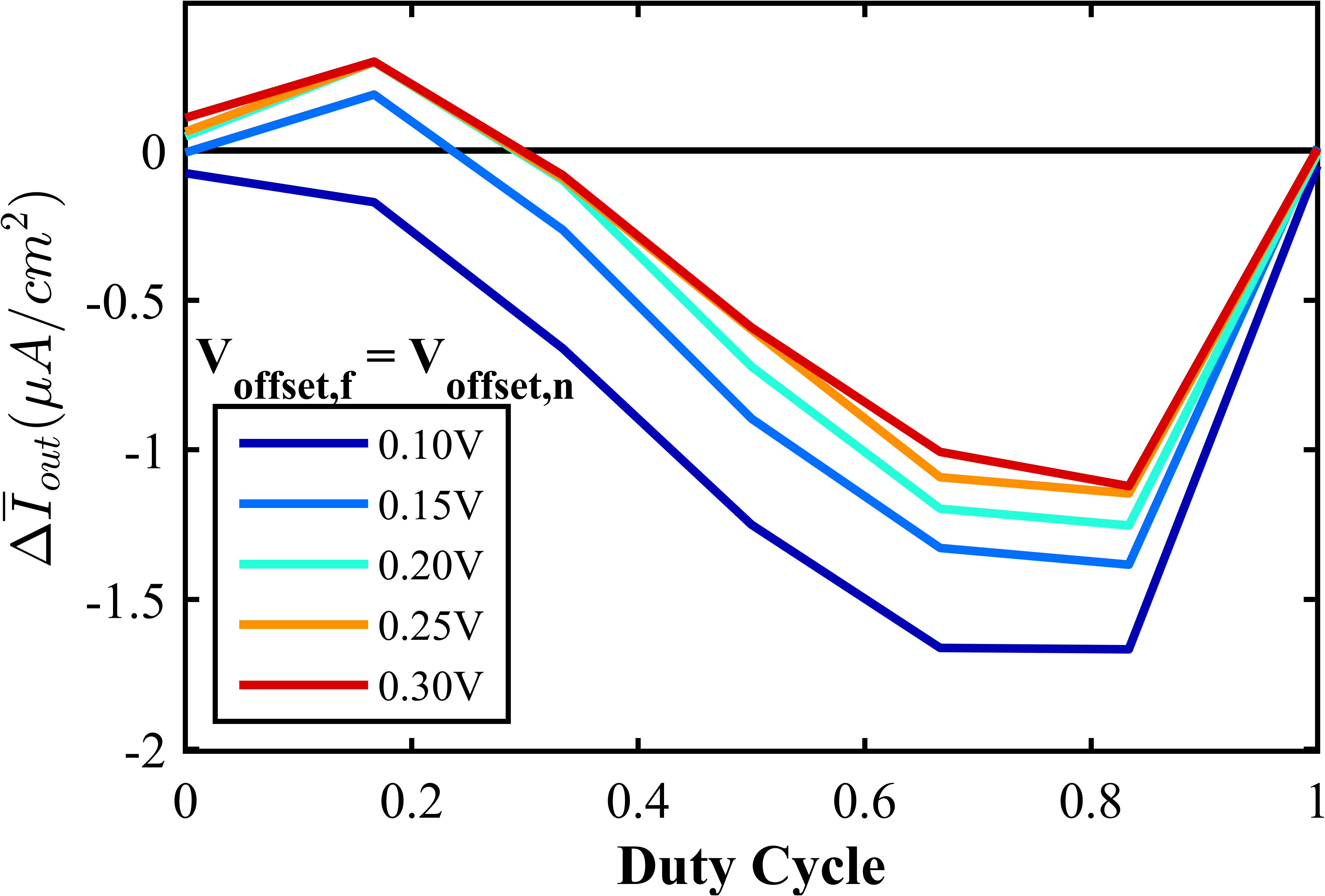}\label{fig:offset,f=offset,n}}
    \hfill
  \subfloat[]{\includegraphics[width=8cm]{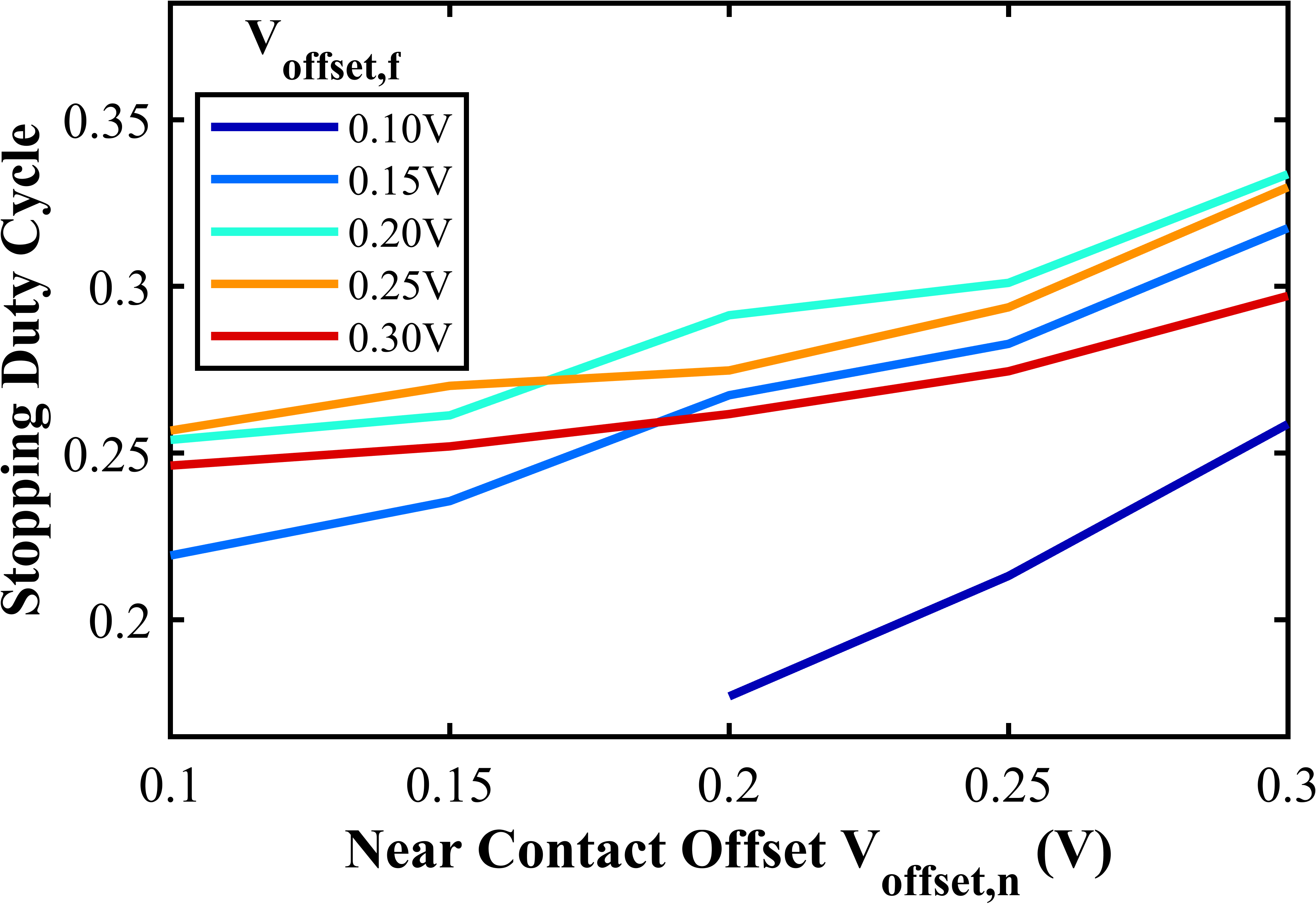}\label{fig:stopping duty cycle}}
    \caption{(a) Output current density as a function of duty cycle for several input signal offsets with $V_{\text{offset,f}} = V_{\text{offset,n}}$.  (b) Stopping duty cycle as a function of $V_{\text{offset,n}}$ and several values of $V_{\text{offset,f}}$.  
    The input signal frequency is 7Hz, and $V_{\text{p-p}} = 0.3$V for both figures.}
\label{fig: offset sweep averages}
\end{figure}

Figure \ref{fig:offset,f=offset,n} shows the output current density as a function of the input signal duty cycle for different voltage offsets with $V_{offset,f}=V_{offset,n}$. The input signal frequency is 7 Hz and the amplitude is $V_{p-p}= 0.3\space  V$. When increasing the offsets, the RBIP characteristic curve shifts from a near-even behavior (no output reversal with duty cycle) to a near-odd behavior (stopping duty cycle near 0.5) \cite{Rozenbaum2019SymmetryRatchets}. 
Figure~\ref{fig:stopping duty cycle} shows the measured stopping duty cycle as a function of $V_{offset,n}$  for several values of $V_{offset,f}$. All other parameters as in Figure \ref{fig:offset,f=offset,n}. The stopping duty cycle increases with $V_{offset,n}$ for all tested values of $V_{offset,f}$. However, the effect of $V_{offset,f}$ on the stopping duty cycle is much less pronounced.
Mirroring the electrode arrangement shown in Figure \figref[b]{fig: simple system} and placing the reference and counter electrodes in the opposite compartment produced the same result - the voltage offset of the contact facing the counter and reference electrodes is more dominant in defining the stopping duty cycle of the device (see Appendix~\ref{SI: offset sweeps}). Similarly, the effect of the applied bias on the stopping duty cycle was observed also when only the potential of the far contact was alternating, and the near contact was held under different constant bias values (Figure~\ref{SI fig: offset sweep when oscillating single contact}).
The offset applied to the near contact controls its potential with respect to the bulk solution next to it. Thus, it sets a well-defined working point despite uncertainty in the potential of zero charge. This defines the two potentials between which the near contact alternates and thus an effective amplitude asymmetry coefficient  that can be tuned with $V_{offset,n}$ (analogous to $\alpha$ in theoretical models). In contrast, the electrostatic potential of the far contact is not set relative to the potential of the bulk solution near it. As a result, $V_{offset,f}$ is less effective at controlling the potential difference between the far contact and the bulk solution next to it, leading to a weaker effect on the asymmetry of the system. Since the amplitude asymmetry coefficient has a significant effect on the RBIP separation properties \cite{Herman2023Ratchet-BasedSeparations}, tuning this asymmetry with the application of a constant bias, $V_{offset,n}$, is an important step towards demonstrating ratchet-driven selective ion separation.

\subsection{Transport Mechanism} \label{Ion Pumping}
To better understand the RBIP operating mechanism, the output voltage of a different sample was measured in two configurations: the complementary input signal configuration in which the potential of the contacts was alternated as in Figure \figref[a]{fig:Parameter_sweep}, and when alternating the potential at the far contact while holding the potential of the near contact at 0 V vs reference:
\begin{equation}\label{eq:vin2_definitions}
\begin{aligned}
V_{in,f} =\left\{\begin{array}{l}
-0.05V ,\ 0<t<d_ct  \\
\phantom{-}0.45V ,\ d_ct<t<T
\end{array}
\right. \ \ \ ; \ \ \ 
V_{in,n} =0 V
\end{aligned}
\end{equation}
$V_L$, the voltage between the far RBIP contact and the Ag/AgCl wire next to it, was obtained in both configurations by subtracting the output voltage from the input voltage of the far contact (Figure \figref[a]{fig: mechanism}).
Figure  \figref[b]{fig: mechanism} shows the output voltage for both input configurations. The device consistently generated higher outputs when operated with the complementary input signal. The non-zero current is measured at extreme duty cycles ($d_c=0,1$) due to transient effects (see Figure~\ref{SI fig: stabilization effect} for details).
\begin{figure*}[htbp!]
\includegraphics[width=17.2cm]{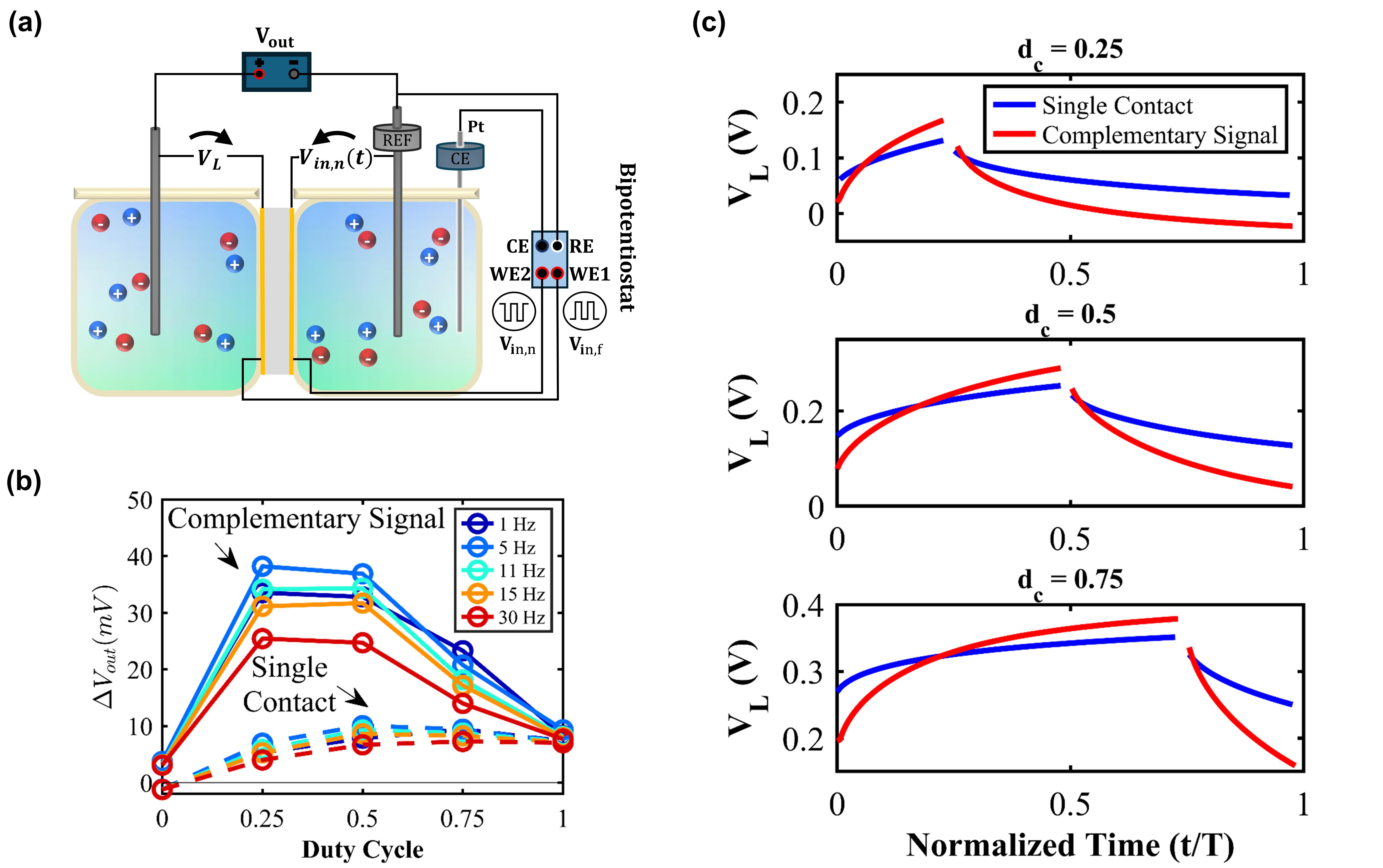}
\caption{(a) The output voltage as a function of the input signal duty cycle when the RBIP is driven with complementary signals (solid lines) and when the potential of only one surface is alternated (dashed lines). (b) Illustration of the measured voltage $V_L$.
    (c) The $V_L(t)$ waveform obtained by averaging the response of 30 temporal periods when driven with complementary input signals and when the potential of only one surface was alternated. The input signal frequency is $f = 5$~Hz.}
\label{fig: mechanism}
\end{figure*}

Figure \figref[c]{fig: mechanism} shows the waveform $V_L(t)$ in the complementary configuration and single-contact input configurations. Each waveform was obtained by averaging the responses of 30  consecutive periods. The standard deviation of the responses was below 1 mV at every point along the period. When a complementary input signal is applied, $V_L$ alternates with significantly higher amplitude. Since there are no electrochemical reactions on the far contact, $V_L$ is comprised mostly of the potential drop across the double layer at the far contact electrode. Thus, a larger amplitude of $V_L$ implies that a higher local charge density accumulates near the far contact when complementary signals are applied. Given that the electrical conditions within the far contact compartment is identical for both input configurations, this increase originates from the action of the near contact. This can be a result of a "push-pull" mechanism facilitating directed ion transport from the near contact to the far contact.

While the two configurations differ in the potential offset of the contacts, the substantially larger output voltage amplitude in the complementary case cannot be explained by offset differences alone. Numerical simulations are required to verify the proposed transient "push-pull" mechanism and to rigorously determine how the alternating potential at the near contact enhances charge transfer through the device. 

The slow evolution of $V_L(t)$ shown in Figure \figref[d]{fig:Parameter_sweep}, defines the spatial asymmetry of the system. Unlike the slow response of $V_L(t)$,  the potential of the near contact with respect to the bulk solution next to it alternates rapidly following $V_{in,n}(t)$. The difference between $V_L(t)$ and $V_{in,n}(t)$  breaks the anti-symmetry and as a result the RBIP output deviates from the odd behavior that can be expected under the complementary drive \cite{Rozenbaum2019SymmetryRatchets}. Thus, while in the floating-drive configuration, device asymmetry originates from inherent differences between the RBIP surfaces \cite{Kautz2025ASeparations}, in the bipotentiostatic drive asymmetry is a result of externally imposed electrical conditions, making it less sensitive to surface properties. This low sensitivity was experimentally confirmed by interchanging the reference and counter electrodes between compartments, which resulted in negligible changes in RBIP output (see Appendix \ref{electrodes placement} for more details).

Due to sample degradation, different samples were used in different experiments. Nevertheless, the consistency of qualitative behavior and trends across these samples supports the robustness of our findings. Appendix \ref{Sample performance comparison} provides a comparison of the normalized outputs of all samples. More details on the sample degradation mechanisms can be found in prior work \cite{Kautz2025ASeparations}. Further details on electrode placement effects, working range determination, and potential ionic current contributions from the bipotentiostat are provided in the Appendix ~\ref{BiPotentiostat effcts}.
\section{Conclusions}

The operation of RBIPs under bipotentiostatic control was studied and analyzed. Analogous to flashing ratchets, which employ temporal modulation of the spatial potential distribution, the bipotentiostatic control enables precise and independent modulation of the electrostatic potential at both device contacts.  Implementing a complementary drive scheme resulted in a performance enhancement of up to 1000\%  compared to previous control strategies. Under bipotentiostatic operation, RBIPs exhibited the hallmark characteristics of flashing ratchets, including frequency-dependent reversal of the output current and voltage. Furthermore, the output symmetry was tuned by adjusting the potential offsets of the RBIP contacts. The demonstrated flashing ratchet-like functionality constitutes a significant advancement toward the development of ratchet-driven selective ion separation systems.

\section{Acknowledgments}
This work is partially funded by the European Union (ERC, ESIP-RM, 101039804). Views and opinions expressed are however those of the author(s) only and do not necessarily reflect those of the European
Union or the European Research Council Executive Agency. Neither the European Union
nor the granting authority can be held responsible for them.    

\bibliography{references.bib}

\newpage

\renewcommand\theequation{\Alph{section}\arabic{equation}} 
\counterwithin*{equation}{section} 
\renewcommand\thefigure{\Alph{section}\arabic{figure}} 
\counterwithin*{figure}{section} 
\renewcommand\thetable{\Alph{section}\arabic{table}} 
\counterwithin*{table}{section} 

\newpage
\appendix{Supporting Information}

\section{Detailed description of the experimental setup}
\subsection{Electrical configuration}\label{Electrical configuration}
\begin{figure}[htbp!]
    \centering
    \includegraphics[width=8cm]{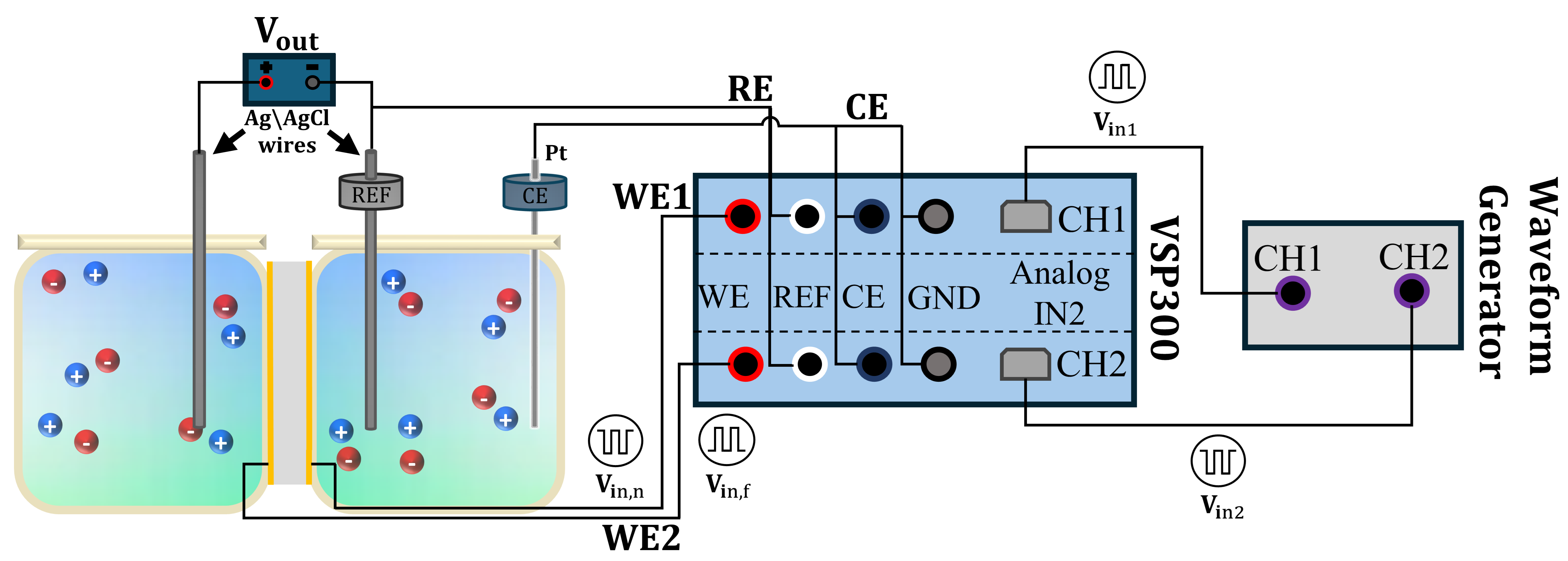}
    \caption{A detailed electrical illustration of the experimental setup}
    \label{SI fig: full system}
\end{figure}

Figure~\ref{SI fig: full system} shows a detailed illustration of the experimental setup. The system consisted of a BioLogic VSP-300 potentiostat with two channels, each equipped with an Ultra-Low-Current (ULC) booster. Each working electrode lead was electrically connected to one of the RBIP gold surfaces. Both reference electrode leads were connected to a single Ag/AgCl wire immersed in one of the solution compartments. The counter electrode and ground leads of both channels were connected to a single platinum wire in the same compartment as the reference electrode. Both channels operated in grounded counter electrode mode.

The two working electrodes potentials (vs. reference) were configured to follow the output voltage of two  Keysight 33522B waveform generators. This was obtained by connecting the output of each waveform generator to the Analog IN2 terminal of each bipotentiostat channel.
The bipotentiostat reliably reproduced square waves for frequencies between 0.1 Hz and 5 kHz.

\subsection{RBIP performance characterization} \label{RBIP performance characterization}
The output current or voltage were measured using a Keysight 34465A digital multimeter. For low-frequency signals (1–30\,Hz), an integration time of 1\,ms enabled time-resolved acquisition of the output signals, from which the average output was subsequently computed. For higher-frequency signals, an integration time of 2\,s was employed, such that the measured values directly reflected the time-averaged response. 

\begin{figure}[htbp!]
    \centering
    \includegraphics[width=8cm]{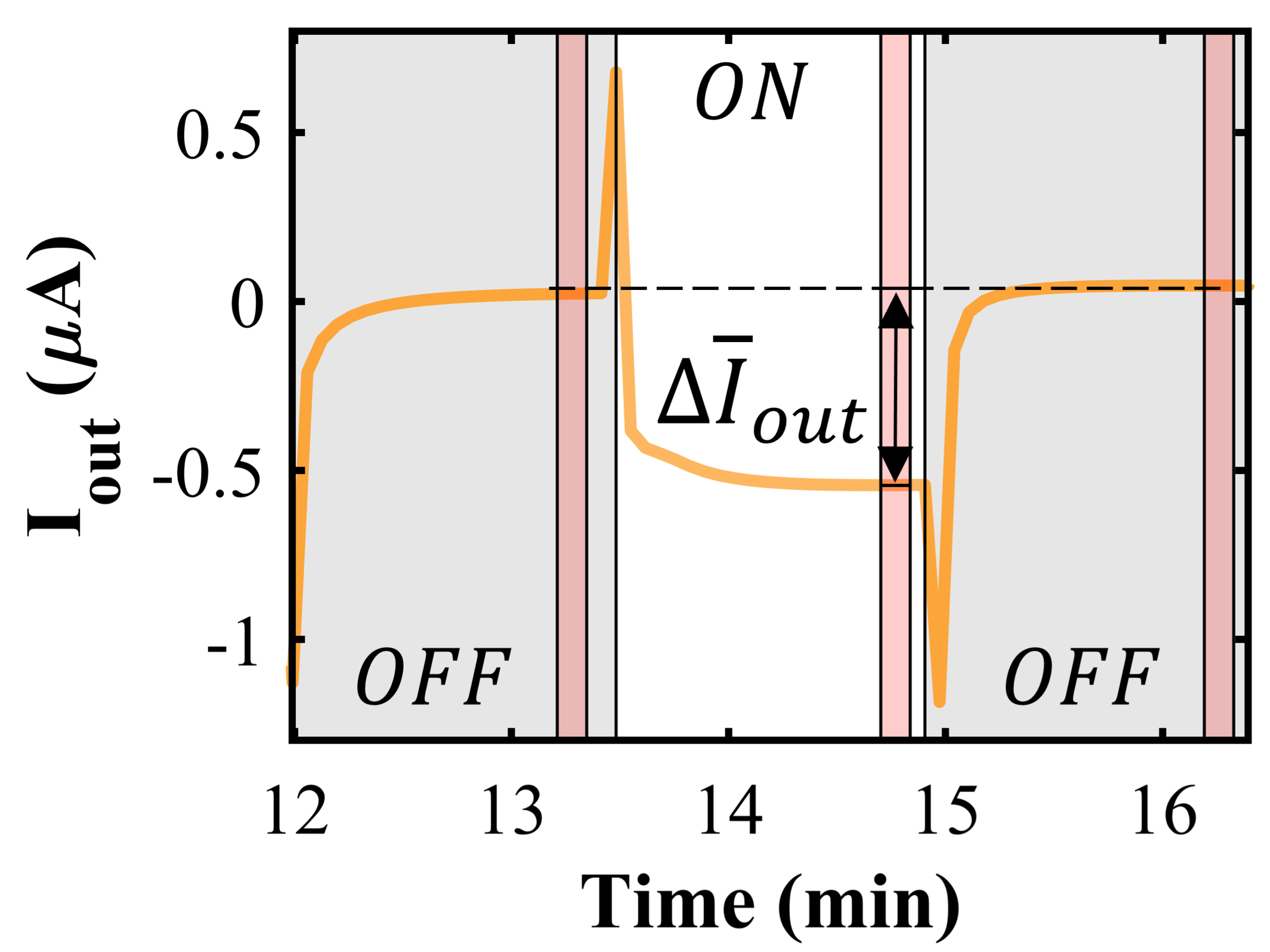}
    \caption{Example of net output current calculation, $\Delta\overline{I}_{\text{out}}$. Gray regions indicate "OFF" segments ($V_{\text{in,f}} = V_{\text{in,n}} = 0\,\text{V}$), white region indicates the "ON" segment, and red regions mark the time between 0.85 and 0.95 of the segments which were used for obtaining the time averaged currents. The net current is obtained by subtracting the mean of the two ratchet OFF time averaged currents from the ratchet ON time averaged current. The input signal frequency is 100\,Hz and the duty cycle is 0.5. }
    \label{SI fig: output definition}
\end{figure}

Most experiments involved parameter scans, primarily of the input signal duty cycle. For each input signal parameters set, the RBIP operated for between 60 s and 120 s ("on" time), followed by an "off" time during which both inputs were set to $V_{\text{in,f}} = V_{\text{in,n}} = 0\text{ V}$ to allow for system relaxation and account for slow drift process. Three mean signal values were calculated: the temporal average of the signal measured between  0.85 and 0.95 of the "ratchet ON" segment, the temporal average of the signal measured between  0.85 and 0.95 of the "ratchet OFF" segment prior to the ratchet operation, and the temporal average of the same part of the "ratchet OFF" segment measured after the ratchet operation. The net output was calculated by subtracting the averaged values obtained from the two "OFF" segments means from the averaged value obtained from the "ON" segment mean. Figure~\ref{SI fig: output definition} shows an example for the net output current calculation, $\Delta\overline{I}_{\text{out}}$. Gray shaded regions indicate segments when the RBIP was OFF, while the white-background region indicates the ON segment. Transients in the output are visible when switching between states, followed by a nearly constant signal. Red shaded regions indicate the time between 0.85 and 0.95 of the segments which was used for calculating the output.

\section{Generality of Results} \label{Sample performance comparison}

\begin{figure}[htbp!]
    \centering
    \includegraphics[width=8cm]{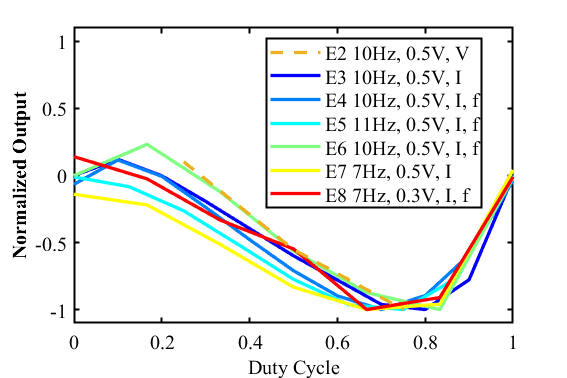}
    \caption{The normalized output of samples E2 through E8 as a function of the input signal duty cycle. The output of each sample was normalized by its absolute maximum. The input signal frequency is between from 7 to 11 Hz and the amplitude was 0.3 or 0.5 V (peak-to-peak). The input signal frequency and amplitude of each experiment  are indicated in the legend as well as the  measurement type (voltage 'V', or current 'I').  Normalized output curves noted with 'f'  were mirrored vertically and horizontally about a duty cycle of 0.5 ('f'). $V_{offset,f} = V_{offset,n} = 0.2$ V in all cases.}
    \label{SI fig: sample comparison}
\end{figure}

Due to sample degradation, the performance characteristics of our RBIPs had to be tested using several different samples. However, the qualitative performance of nearly all samples tested was very similar. Figure~\ref{SI fig: sample comparison} shows the normalized output as a function of duty cycle for all samples discussed in this work. The output of each sample was normalized by dividing it by its own maximum value. The input signal frequency of all samples is between 7 and 11 Hz. Since the CV curves of the samples are linear, the normalized outputs allow a simple comparison of both current and voltage measurements. To compare samples measured with different electrode configurations, some normalized output curves were mirrored vertically and horizontally about a duty cycle of 0.5.

While the absolute output varied between samples, the normalized outputs reveal a consistent trend across all devices tested. This consistency implies that the observed ratchet features represent general device characteristics rather than sample-specific artifacts thus supporting the generalizability of the reported findings.

\section{Bipotentiostat effects} \label{BiPotentiostat effcts}
\subsection{Working range} \label{Working range}

To ensure all experiments operated within an electrochemically stable potential window, cyclic voltammetry (CV) measurements were conducted using a three-electrode configuration. In this setup, each RBIP contact (gold working electrode) was tested individually against the Ag/AgCl wire quasi-reference electrode with a platinum counter electrode, as illustrated in Figure \ref{fig: simple system}. During each measurement, one RBIP surface served as the working electrode while the other was disconnected, allowing independent electrochemical characterization of  each surface.

\begin{figure}[htbp!]
    \centering
    \subfloat[]{\includegraphics[width=0.5\textwidth]{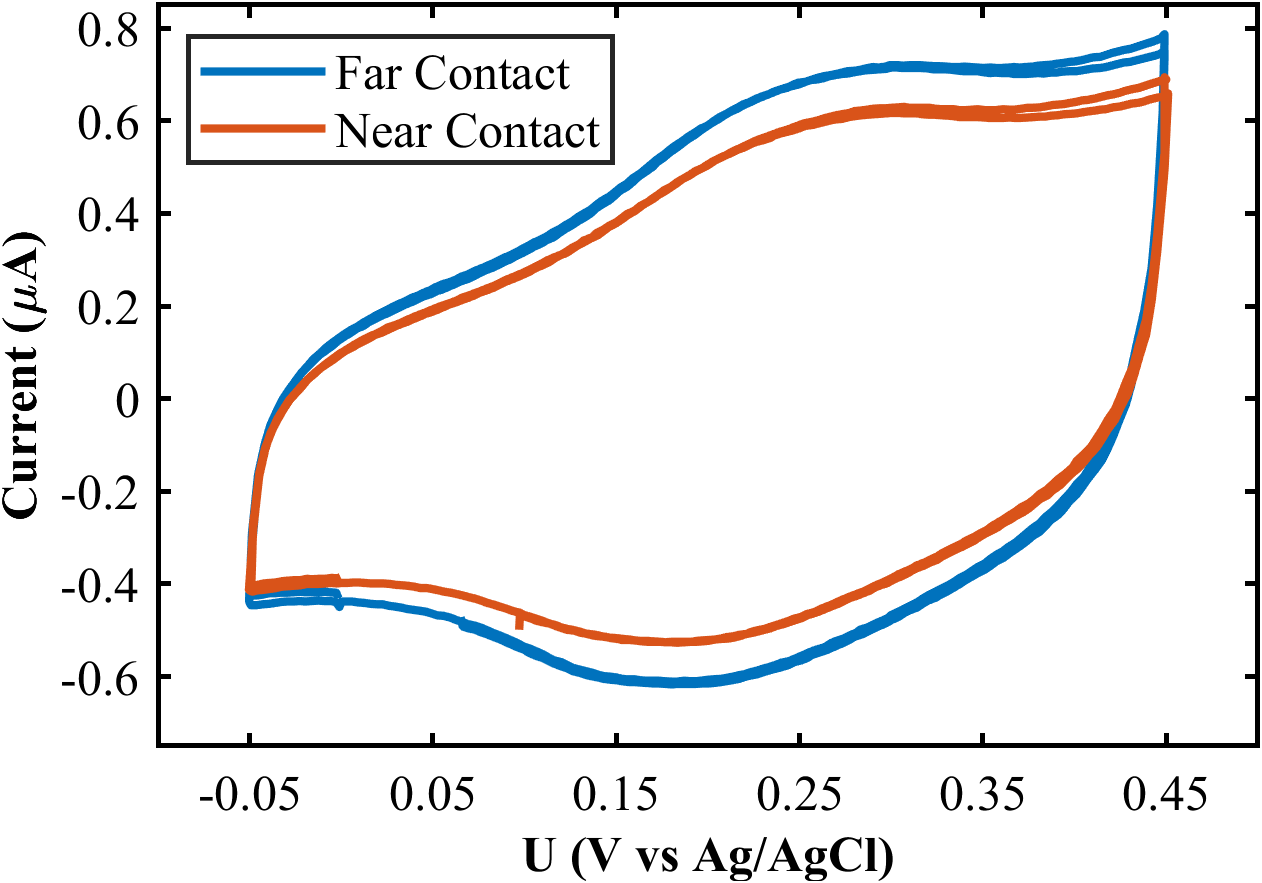}\label{fig: CV vs ref}}
    \hfill
       \subfloat[]{\includegraphics[width=0.5\textwidth]{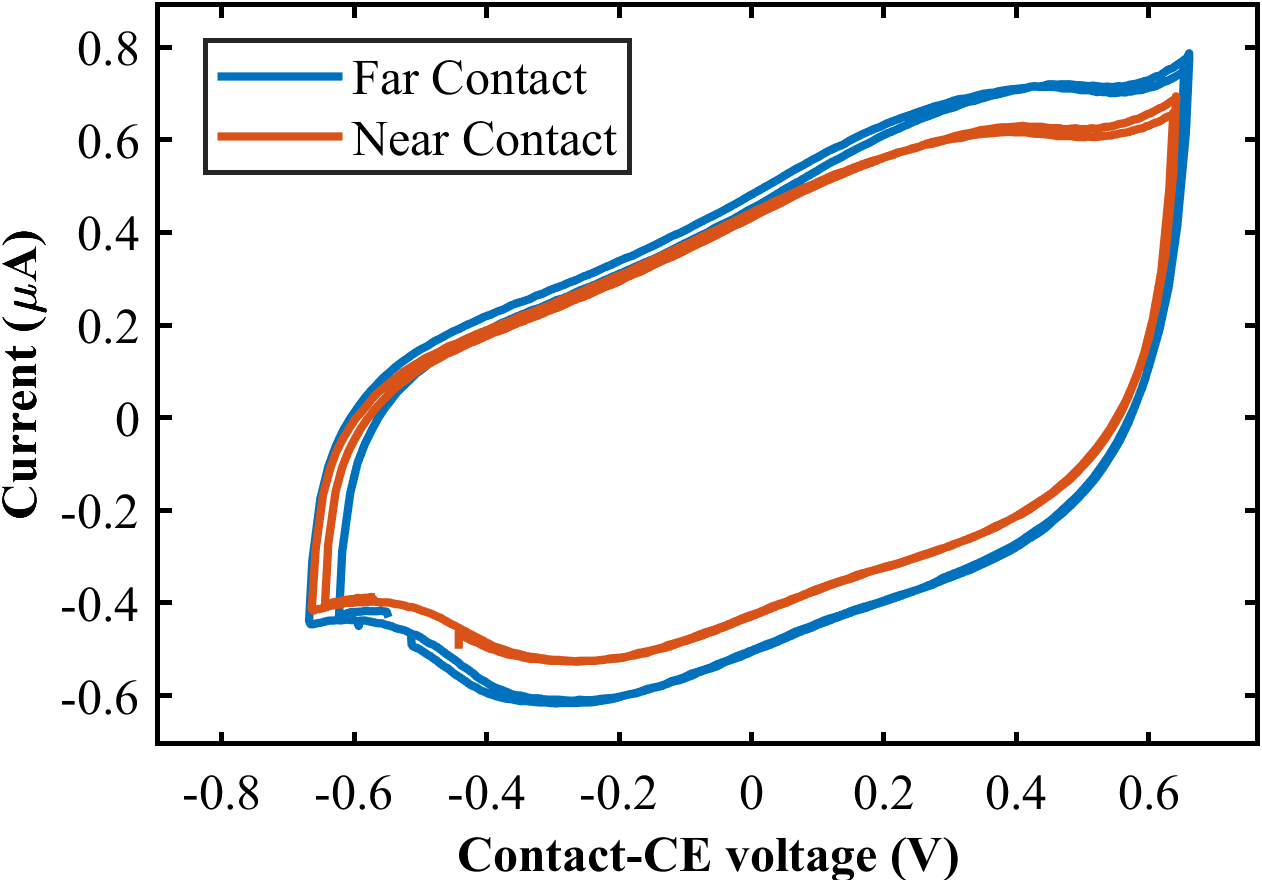}\label{fig: CV vs CE}}
    \caption{Cyclic voltammetry measurements of both RBIP surfaces in a three-electrode configuration using an Ag/AgCl wire quasi-reference electrode and platinum counter electrode. Each surface was tested independently as the working electrode while the other was disconnected. Scans were performed between of –0.05 V to 0.45 V vs. Ag/AgCl at a scan rate of 10 mV/s. (a) shows the  voltammograms measured between each surface and the reference electrode; (b) shows the same measured current as a function of the voltage between each surface and the counter electrode.}
    \label{SI fig: CV}
\end{figure}

Figure \ref{fig: CV vs ref} shows representative CV results for both surfaces. The potential was scanned between  -0.05 V and 0.45 V vs. Ag/AgCl at a scan rate of 10 mV/s. No Faradaic reactions were observed within this range, as evidenced by the absence of redox peaks in the voltammograms.  Figure \ref{fig: CV vs CE} shows the same measurements as a function of the voltage between each surface and the counter electrode. In the entire potential range tested, the voltage between working and counter electrodes remained well below the threshold for water electrolysis (1.23 V), ensuring purely capacitive operation without unwanted electrochemical side reactions. Based on these measurements, all experiments employed various signal amplitudes and offsets that remained confined within this established non-Faradaic working range of -0.05 V to 0.45 V vs. Ag/AgCl, thus ensuring capacitive operation across all experimental conditions.

\subsection{System directionality}  \label{electrodes placement}
\begin{figure}[htbp!]
    \centering
    \includegraphics[width=8cm]{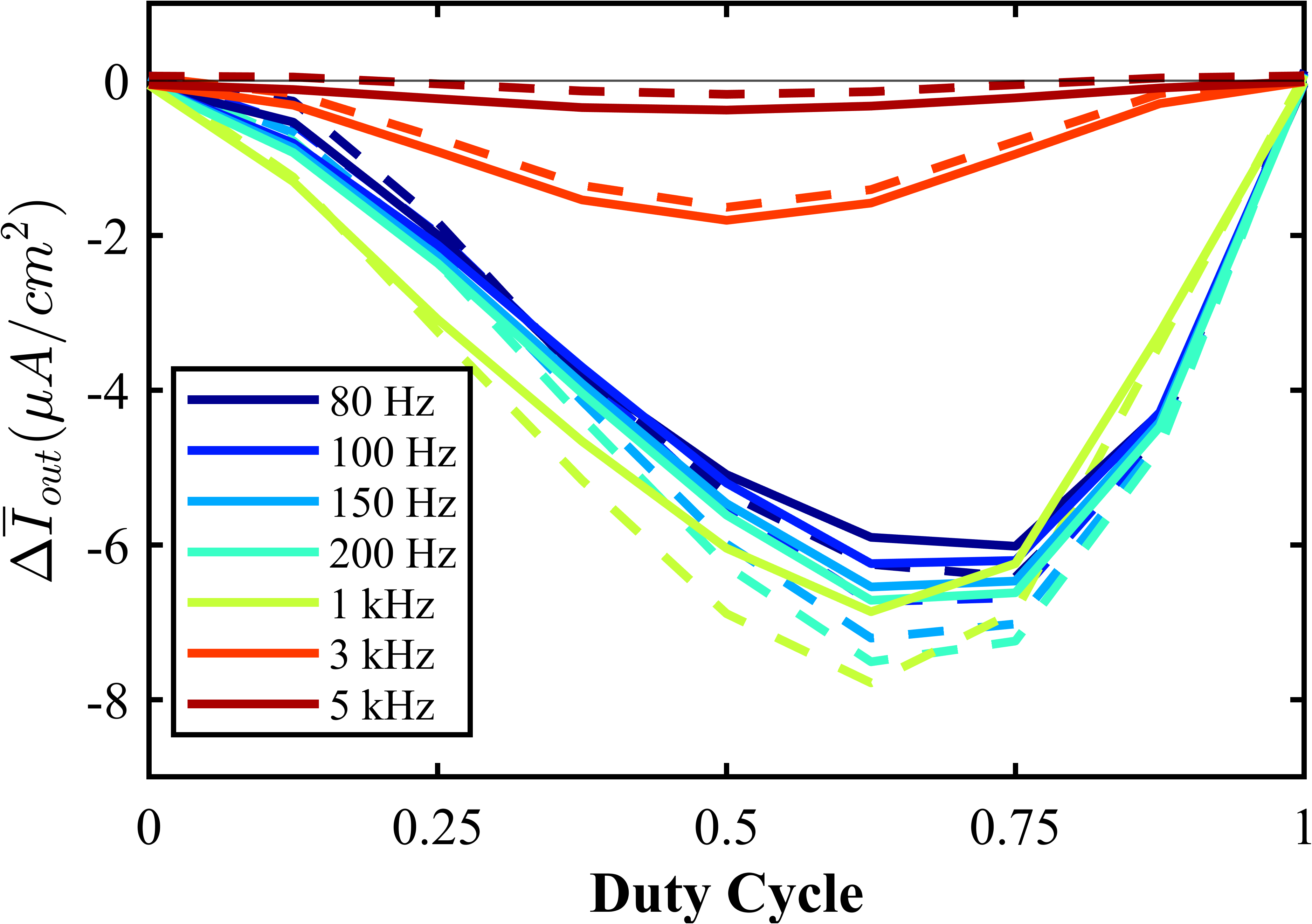}
    \caption{The output current as a function of the duty cycle for several input signal frequencies. Solid lines: the reference and counter electrodes were placed in one compartment (Figure \figref[b]{fig: simple system}); dashed lines: the counter and reference electrodes were placed in the opposite compartment. The duty cycle was defined as that of the same physical device contact, and the direction of current measurement remained unchanged between configurations.}
    \label{SI fig: CE1 vs VE2}
\end{figure}

Figure~\ref{SI fig: CE1 vs VE2} shows the output current as a function of the duty cycle when the counter and reference electrodes were positioned in one compartment, as illustrated in Figure \figref[b]{fig: simple system} (solid line), and when the counter and reference electrodes were placed in the opposite compartments (dashed lines). The RBIP performance is independent of the compartment which contains the reference and counter electrodes (as long as $V_{offset,n}=V_{offset,f}$). The observed variation between different electrode configurations is comparable to the variation between repeated measurements of the same configuration and is attributed to gradual device degradation during testing. These results confirm that the inherent asymmetry observed in device measurements is not determined by which compartment houses the reference and counter electrodes of the membrane itself. Instead, the applied voltage signals defines the asymmetry, thereby allowing a simple and repeatable control of the ion pumping properties.

\section{Voltage Offset effects}
\label{SI: offset sweeps}
Figure~\ref{fig:stopping duty cycle} shows that the stopping duty cycle and thus the device asymmetry can be tuned by applying a potential offset to the RBIP contact facing the reference and counter electrodes. To verify whether this is due to the bipotentiostatic drive or structural asymmetry in the sample, the experiment was repeated with the counter and reference electrodes placed in the opposite compartment. The potential offset of the contact facing the reference and counter electrodes in the new configuration is noted $V_{offset,n}^*$ and the potential offset of the opposite contact is noted $V_{offset,f}^*$. The duty cycle definition and current polarity are consistent with the original configuration.

\begin{figure}[htbp!]
  \centering
    {\includegraphics[width=8cm]{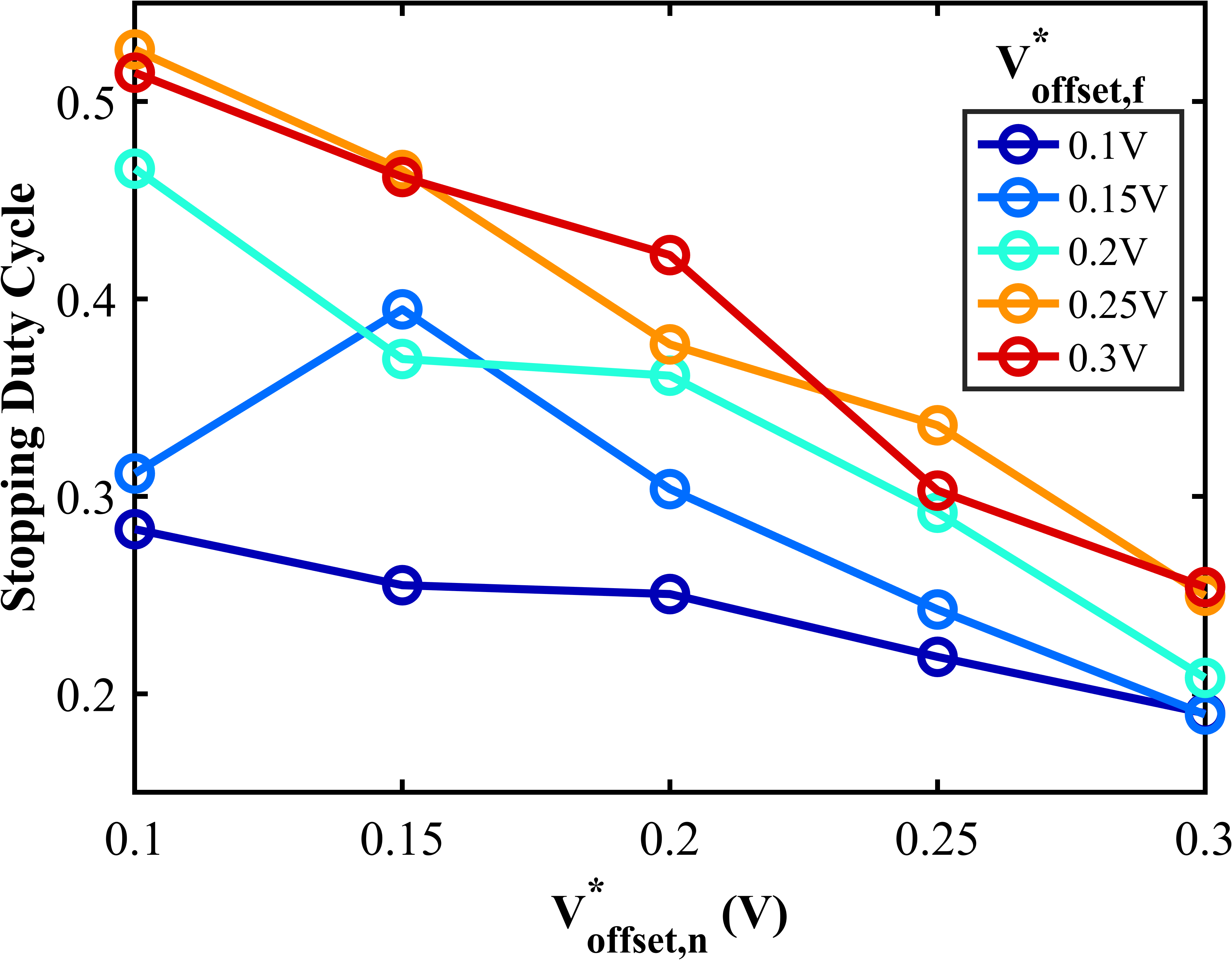}}
      \caption{The stopping duty cycle as a function of the contact potential offset $V_{offset,n}^*$ for several values of $V_{offset,f}^*$. The sample as in  Figure~\ref{fig:stopping duty cycle}, the input signal frequency is $f = 7$ Hz, and the amplitude is  $V_{p-p} = 0.3$ V.}

\label{SI fig: offset sweep when moving the electrodes}\end{figure}

Figure~\ref{SI fig: offset sweep when moving the electrodes} shows the measured stopping duty cycle as a function of $V_{offset,n}^*$ for several values of $V_{offset,f}^*$ in the new configuration.  Increasing  $V_{offset,n}^*$ leads to a near-monotonic decrease in the stopping duty cycle. However,  increasing $V_{offset,f}^*$ results in a less pronounced change in stopping duty cycle, especially at higher values of $V_{offset,n}^*$. 

\begin{figure}[htbp!]
  \centering
    \subfloat[]{\includegraphics[width=8cm]{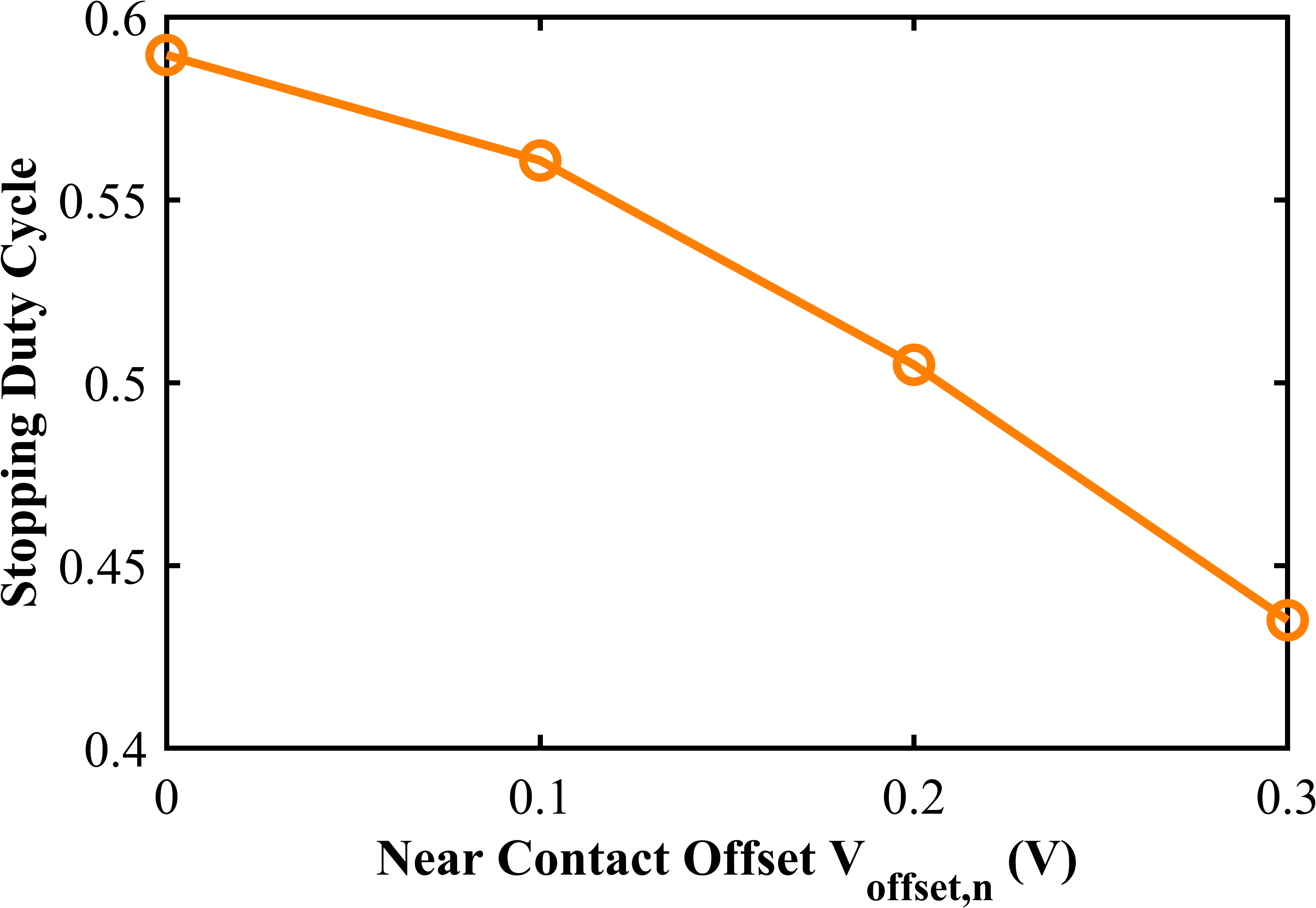} \label{fig: E6 - CE@2 offset sweep osc only far}}
    \hfill
  \subfloat[]{\includegraphics[width=8cm]{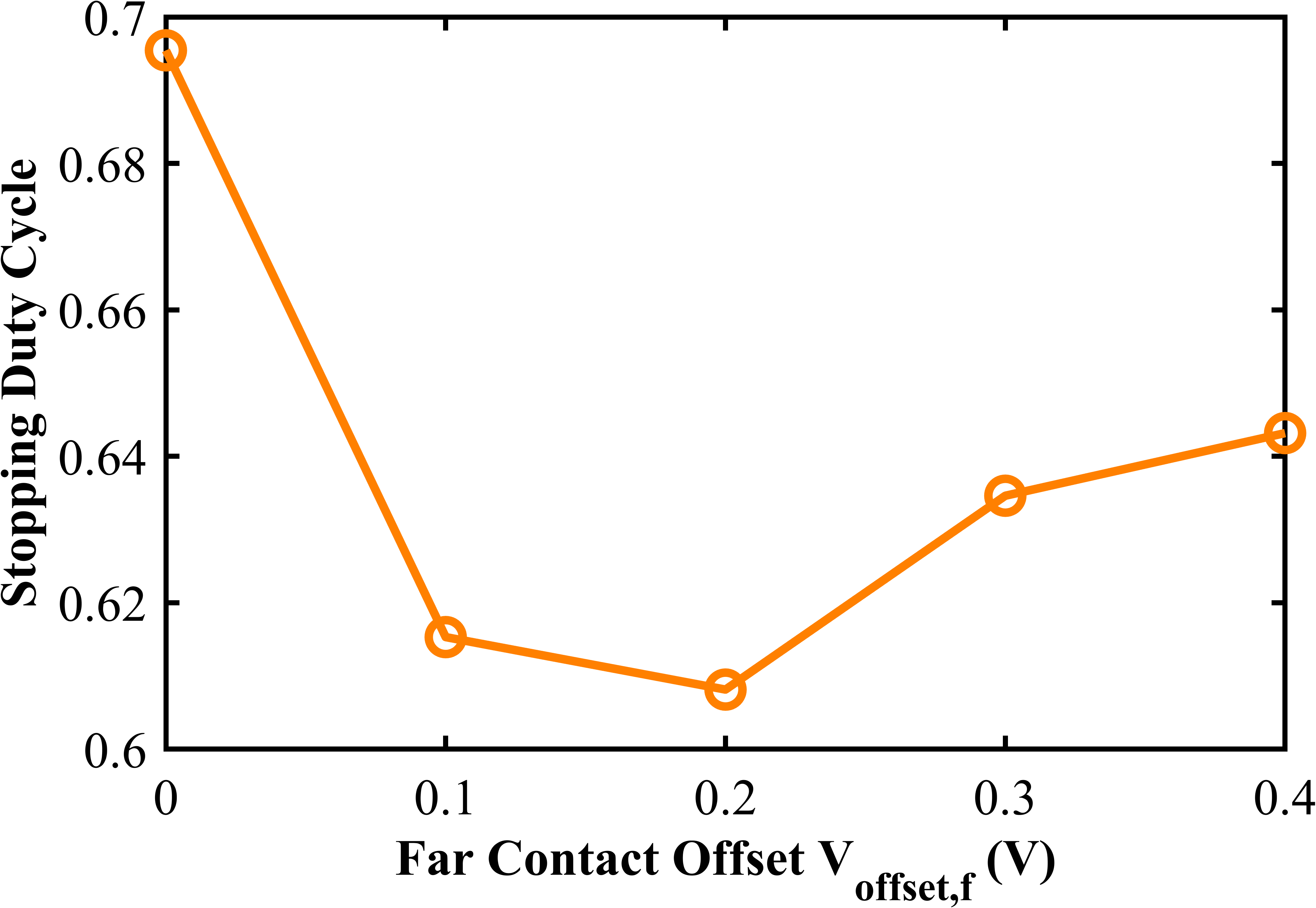}\label{fig: E6 - CE@2 offset sweep osc only close}}
    \caption{The stopping duty cycle when alternating the potential of a single contact while varying the potential offset of the other contact ($f = 100$ Hz, $V_{p-p} = 0.5$ V). (a) Altarnating the potential of the far contact while the near contact offset is varied. (b) Alternating the potential of the near contact while while applying a constant bias to the far contact offset is varied.}
\label{SI fig: offset sweep when oscillating single contact}\end{figure}

The change in the stopping duty cycle with potential offset was also studied when the potential of only one contact was alternating. Figures~\ref{fig: E6 - CE@2 offset sweep osc only far} and \ref{fig: E6 - CE@2 offset sweep osc only close} show the stopping duty cycle of  a second sample when alternating the potential of one contact while the potential of the other was not alternated but held at a constant bias. When the far contact alternated and the near contact remained at constant offset (Figure~\ref{fig: E6 - CE@2 offset sweep osc only far}), varying the near contact offset shifted the stopping duty cycle substantially. However, when the potential of the near contact was alternated and the far contact was held at a constant bias (Figure \ref{fig: E6 - CE@2 offset sweep osc only close}), changing the far contact offset did not lead to a systematic change in the stopping duty cycle. These results further show that near-contact offset is more dominant in determining stopping duty cycle and system asymmetry.

\section{Transient effects}
\begin{figure}[htbp!]
    \centering
    \includegraphics[width=0.6\linewidth]{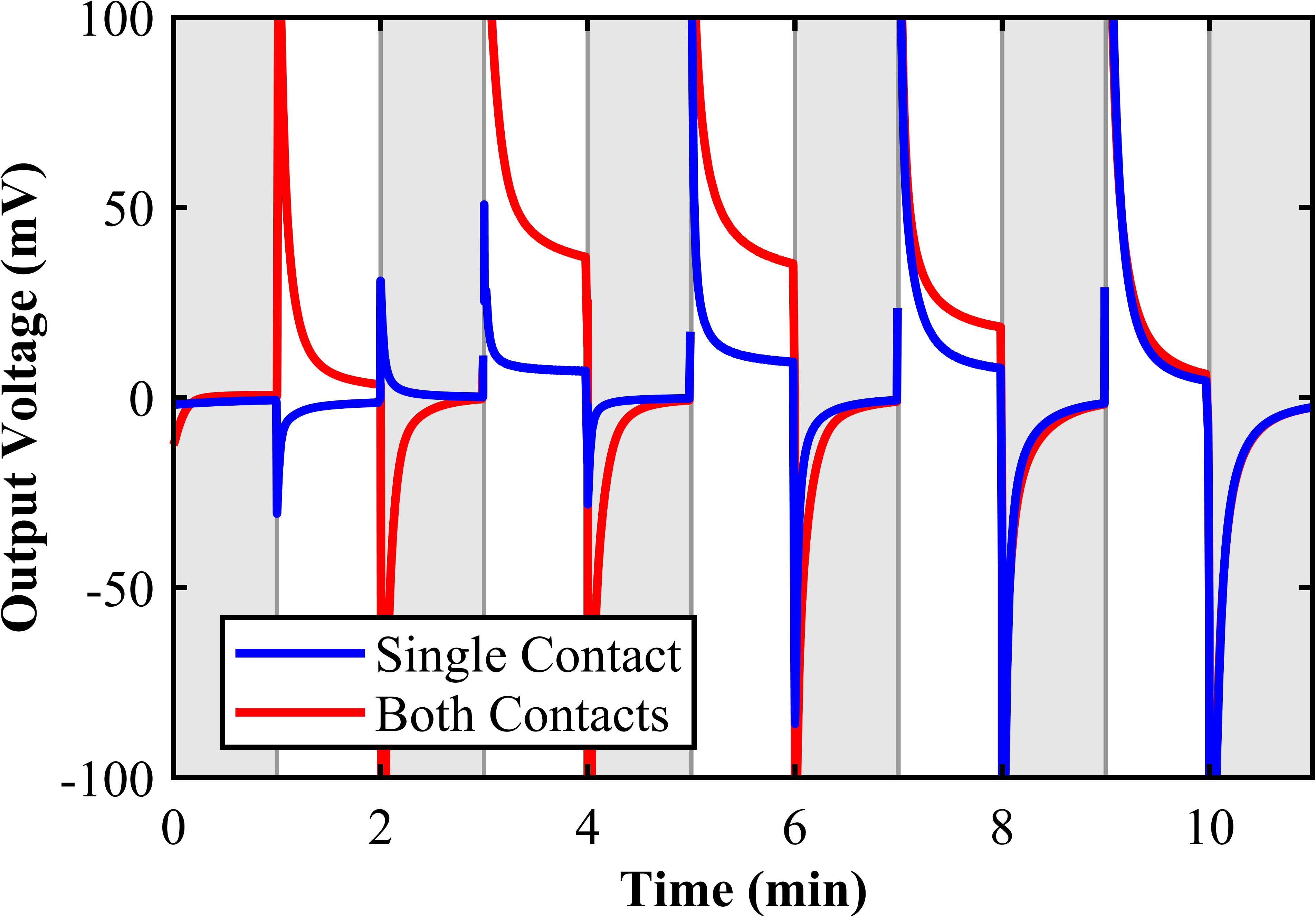}
    \caption{Output voltage sweep at $f = 5$ Hz when oscillating a single contact or both, as defined in Section~\ref{Ion Pumping}. Data were sampled at $f_s = 833.33$ Hz, and a moving average over 1000 points (1.2 s) was applied. White background regions indicate periods when the ratchet was active, corresponding to sequential duty cycles of 0, 0.25, 0.5, 0.75, and 1. Gray background regions indicate periods when the device was OFF, with inputs $V_{in,f}=V_{in,n}=0$ V.}
    \label{SI fig: stabilization effect}
\end{figure}

As seen in Figures~\ref{fig:offset,f=offset,n} and \figref[b]{fig: mechanism}, a small non-zero output was measured even at duty cycles of 0 and 1. This occurred because transient effects induced by changes in the input signal and the corresponding change in equivalent DC voltage do not fully decay before the next signal change. Figure~\ref{SI fig: stabilization effect}  shows the voltage measured over the course of this experiment.  The processed results of which are displayed in Figure \figref[b]{fig: mechanism}. The characteristic exponential decay is clearly visible; however, the relaxation time exceeds the on periods at these extreme duty cycles, preventing complete decay and causing the observed residual output.

\end{document}